\documentclass[12pt]{article}
 \def\cen{\centerline}

\begin{document}

\setlength{\unitlength}{1mm}

 \title{Bianchi -I  Cosmology with Magnetic Field  in Lyra Geometry.}
 \author{\Large $F.Rahaman^*$,N.Begum,S.Das ,M.Hossain and M.Kalam}
\date{}
 \maketitle
 \begin{abstract}
         A new class of Bianchi - I  cosmological model with magnetic field within
         the frame work of Lyra geometry has been presented . Exact solutions to
         the field equations for the model are obtained . The physical and kinematical
         behaviors of the model have been discussed.

  \end{abstract}


 \cen{ \bf 1. INTRODUCTION }

 \bigskip
 \medskip
  \footnotetext{ Pacs Nos : 98.80cq ,  04.20jb ,  04.50 \\
     \mbox{} \hspace{.2in} Key words and phrases  : Bianchi - I space time, magnetic field , Lyra geometry.\\
                              $*$Dept.of Mathematics,Jadavpur University,Kolkata-700 032,India\\
                              E-Mail:farook\_rahaman@yahoo.com
                              }

    \mbox{} \hspace{.2in} The aim of cosmology is to determine the large scale structure
    of the Universe. The present day observations indicate that the Universe at large
    scale is homogeneous and isotropy. It is well known that in general theory of relativity
    spatially homogeneous space times belong to either to the Bianchi classifications or to
    Kantowski-Sachs class and interpreted as cosmological model
    [1].\\
    In last few decades there has been considerable interest in alternative theory of
    gravitation. The most important among them being scalar tensor theories proposed
    by Lyra [2] and Brans-Dicke [2].Lyra proposed a modification of Riemannian geometry
    by introducing a gauge function into the structure less manifold that bears a close
    resemblances to Weyl's geometry. In general relativity Einstein succeeded in geometrising
    gravitation by identifying the metric tensor with gravitational potentials.
    In scalar tensor theory of Brans-Dicke on the other hand, the scalar field remains
    alien to the geometry. Lyra's geometry is more in keeping with the spirit of Einstein's
    principle of geometrisation since both the scalar and tensor fields have more or less
    intrinsic geometrical significance.\\
    In consecutive investigations Sen [3] and Sen and Dunn [4] proposed a new scalar tensor
    theory of gravitation and constructed an analog of the Einstein field equation based on  \pagebreak
      Lyra's geometry which in normal gauge may be written as\\

    \begin{equation}
              R_{ab} - \frac{1}{2}g_{ab}R + \frac{3}{2}\phi_a
              \phi_b - \frac{3}{4}g_{ab}\phi_c\phi^c = - 8\pi G
              T_{ab}
           \label{Eq1}
           \end{equation}

          Where $\phi_a$ is the displacement vector and other symbols have their usual meaning
    as in Riemannian geometry.\\
     Halford[Aust.J.Phys. 23, 833, (1970) ] has pointed out that the constant displacement
     field  $\phi_a$ in Lyra's geometry play the role of cosmological constant $\Lambda$ in
     the normal general relativistic treatment. According to Halford  the present theory
     predicts the same effects within observational limits, as far as the classical solar system
     tests are concerned, as well as tests based on the linearised form of field
     equations.\\
     Subsequent investigations were done by several authors in scalar tensor theory and cosmology
     within the frame work of Lyra geometry [5].\\
    It is well known that the magnetic field has a significant role at the cosmological
    scale and is present in galactic and intergalactic spaces . The occurrence of magnetic
    fields on galactic scales  and their importance for a variety of   Astrophysical phenomena
    has been pointed out by several authors[6] . Melvin[M.A.Melvin (1975),Ann.N.Y.Acad.Sci. 262,253]
    in his cosmological solution for dust and electromagnetic field argues that the presence
    of magnetic field is not as unrealistic as it appears to be , because for a large part of
    the history of evolution  matter was highly ionized and matter and field were smoothly
    coupled  . Latter during cooling  as a result of expansion of ions  combined to form
    neutral matter.  The cosmological models with magnetic field have been discussed by a
    number of authors in general relativity . But as far as our knowledge there has not been
    any work in literature where Lyra's geometry has been considered to study cosmological model
    with  magnetic field . So it is interesting to study cosmological model with magnetic field
    within the frame work of Lyra geometry. In the present work , we consider Bianchi -I
    cosmological model with magnetic field based on Lyra
    geometry.\\
     \bigskip

 \cen{ \bf 2. The Basic Equations  }
 \bigskip
    We consider an  axially symmetric  Bianchi - I  metric as \\
  \begin{equation}
              ds^2= - dt^2 + e^{2\alpha} dx^2 + e^{2\beta}(dy^2+dz^2)
         \label{Eq2}
          \end{equation}
                 where $ \alpha=\alpha(t) $ and $ \beta=\beta(t) $ .\\
   Now  the energy momentum tensor for the cosmological model with a magnetic field along the  x - direction  is given by\\
   \cen{$T^a_b = T^a_b(c) + T^a_b(m)$ with} \\
    \begin{equation}
             T_{ab}(c)=(p+\rho)U_a U_b + p g_{ab};
            U_i U^i = -1
         \label{Eq3}
          \end{equation}
Where $ U_i $ is the four velocity ,p is the pressure and
          $\rho $ is mass energy density and \\
    \begin{equation}
             T_{ab}(m)= \frac{1}{4\pi}[ F^c_a  F^b_c -
             \frac{1}{4} F^{cd}_{cd}  \delta^b_a ]
         \label{Eq4}
          \end{equation}
    In the above ,$T_{ab}(c)$ is the stress energy  tensor for  a perfect fluid and
    $F_{ab}$ is the electro magnetic  field tensor .\\
    Further , since the magnetic field is being assumed in the  x - direction ,
    $T_{23}$ is the only  non zero component of the electro magnetic  field tensor
    .\\
    Maxwell equation  $F_{[ab,d]} = 0$ and $ [F^{ab}\sqrt{-g}],_a = 0 $, lead to the result
    \begin{equation}
             T_{23} =  A
         \label{Eq5}
          \end{equation}
    A being a constant quantity . So, the components of the stress energy  tensor for
    the electro magnetic field are :
    \begin{equation}
             E^0_0=E^1_1=-E^2_2=-E^3_3=-\frac{A^2}{8\pi}e^{-4\beta}
         \label{Eq6}
          \end{equation}
      The time like  displacement vector $\phi_i$ in (1)  is given by
      \begin{equation}
              \phi_i = ( \beta_0(t),0,0,0 )
         \label{Eq7}
          \end{equation}
      Now choosing  units  such that $8\pi G = 1$, the field
      equation(1) becomes with \\
      $equations (2) , (3) and (6)$ :
       \begin{equation}
            2 \alpha^\prime \beta^\prime+ (\beta^\prime)^2 = \rho
            +\frac{3}{4}\beta_0^2 + \frac{A^2}{8\pi}e^{-4\beta}
         \label{Eq8}
          \end{equation}
       \begin{equation}
            2  \beta^{\prime\prime}+ 3(\beta^\prime)^2 = - p
            -\frac{3}{4}\beta_0^2 + \frac{A^2}{8\pi}e^{-4\beta}
         \label{Eq9}
          \end{equation}
       \begin{equation}
          \alpha^{\prime\prime}+ (\alpha^\prime)^2 + \beta^{\prime\prime}+ (\beta^\prime)^2 +
          \alpha^\prime \beta^\prime = - p
            -\frac{3}{4}\beta_0^2 - \frac{A^2}{8\pi}e^{-4\beta}
         \label{Eq10}
          \end{equation}
         [$^\prime$ denotes the differentiation w.r.t.   't' ]\\

        We assume the equation of state as \\
        \begin{equation}
              p = m \rho  [ 0 \leq  m  \leq 1 ]
         \label{Eq11}
          \end{equation}

        The proper volume , expansion scalar $\theta$ and shear scalar $\sigma^2 $ are
        respectively given by\\
        \begin{equation}
              R^3  =  e^{(\alpha  + 2 \beta)}
         \label{Eq12}
          \end{equation}
        \begin{equation}
              \theta =    \alpha^\prime  + 2 \beta^\prime
         \label{Eq13}
          \end{equation}
         \begin{equation}
               \sigma^2 = (\alpha^\prime)^2  + 2 (\beta^\prime)^2 -
               \frac{1}{3}\theta^2
         \label{Eq14}
          \end{equation}

 \bigskip
 \medskip

 \cen{ \bf 3.   Solutions: }

 \bigskip
 \medskip

     Since the number of unknown parameters appearing in the model exceeds the number
     of field equations by one , we requires one more equation to find the exact solutions .
     This is achieved by assuming  the following relation  between the metric coefficients

    \begin{equation}
              \alpha = a \beta
         \label{Eq15}
          \end{equation}
    Where $'a'$ is an arbitrary constant .\\
    From the field $ equations  (9) and (10)$ , by using $ eq.(15) $, we get

    \begin{equation}
           \beta^{\prime\prime}+ (a+2)(\beta^\prime)^2 =\frac{A^2}{4(a-1)\pi}e^{-4\beta}
         \label{Eq16}
          \end{equation}
      The first integral of $eq.(16)$  gives :

\begin{equation}
     e^{2(a+2)\beta} (\beta^\prime)^2 = -
     \frac{A^2}{8a(a-1)\pi}e^{2a\beta}+ D
\label{Eq17}
          \end{equation}
               (Where D is an integration constant )\\
  The first integral can be written in the integral form  :
  \begin{equation}
     \int \frac{ e^{2 \beta}d\beta}{ \sqrt{D e^{-2a\beta}-
     \frac{A^2}{4(a-1)\pi }}} = \pm ( t - t_0 )
 \label{Eq18}
          \end{equation}
        ( $t_0 $  is another integration constant )\\

  We shall solve the above integral for three different cases :
  $ a = -1 , - 2  $ and $ -\frac{1}{2}$   .\\
   For other cases  the solutions will not be obtained in closed form of $\beta
   $.\\

\cen{ \bf Case - I :  a =  - 1 : }

   Here we get an expression  of $\beta $ as
\begin{equation}
     e^{2 \beta}  = \frac{D^2 (t - t_0 )^2 + \frac{A^2}{16 \pi}}{D}
\label{Eq19}
          \end{equation}
     The other parameters are obtained as\\
\begin{equation}
     R^3   = \sqrt{\frac{D^2 (t - t_0 )^2 + \frac{A^2}{16\pi}}{D}}
\label{Eq20}
          \end{equation}

\begin{equation}
       \theta =\sqrt{  \frac{D^2}{D^2 (t - t_0 )^2 + \frac{A^2}{16\pi}}
       - \frac{A^2 D^2}{16\pi}\frac{1}{{D^2 (t - t_0 )^2 +
       \frac{A^2}{16\pi}}^2}}
\label{Eq21}
          \end{equation}

\begin{equation}
       \sigma^2 =\frac{8}{3} [ \frac{D^2}{D^2 (t - t_0 )^2 + \frac{A^2}{16\pi}}
       - \frac{A^2 D^2}{16\pi}\frac{1}{{D^2 (t - t_0 )^2 +
       \frac{A^2}{16\pi}}^2}]
\label{Eq22}
          \end{equation}

          \begin{equation}
        \rho = p = 0
\label{Eq23}
          \end{equation}
 \begin{equation}
         \frac{3}{4}\beta_0^2 = \frac{mD^2/(1-m)}{D^2 (t - t_0 )^2 +
         \frac{A^2}{16\pi}} - \frac{(2-m)A^2D^2/16\pi(1-m)}{[D^2 (t - t_0 )^2 +
         \frac{A^2}{16\pi}]^2}
\label{Eq24}
          \end{equation}

\bigskip
 \medskip

\cen{ \bf Behavior of the model : }

 \medskip

In this case the solutions degenerate into singularity-free
vacuum solution  based on Lyra geometry. The space time of the
class of solution is to represent an expanding Universe for $t > 0.$\\
 The Kinematical variables $\theta $ and $\sigma^2$ will be vanished as $ t \rightarrow \infty
 .$
 We also see that the ratio $\sigma $ to $\theta $  is constant . This implies that there is
 no possibility that  the model may got isotropized  in some latter  time i.e. it remain
 anisotropic  for all  time .\\

   In this  model particle horizon exists because \\

\cen{ $        \int_{t_0}^t \sqrt[ 6 ]{\frac{D}{D^2 (t^\prime -
t_0 )^2 + \frac{A^2}{16\pi}}} dt^\prime $}

               is a convergent integral .

    The gauge function decreases with the evolution of the model and at $t \rightarrow \infty
     , \beta \rightarrow  0 .$
   So the concept of Lyra geometry will not linger for infinite time
   .\\

   \cen{ \bf Case - II : a = - 2 : }

 \medskip

   For this case integral (18) can at once be evaluated  to yield

          \begin{equation}
         e^{2\beta} = \frac{A}{\sqrt{48 \pi D}}\cosh
         2\sqrt{D(t-t_0)}
         \label{Eq25}
          \end{equation}
    Here we get an expression of $\rho $ as

     \begin{equation}
          \rho = -\frac{4D}{(1-m)\cosh^2  2\sqrt{D(t-t_0)}}
         \label{Eq26}
          \end{equation}

       Since $ m \leq 1 $, the model violets the positivity energy condition .
       So we reject  this case.\\

\cen{ \bf Case - III : a = -$ \frac{1}{2}$  :  }

 \medskip

     We can now find a solution of the integral (18) as

 \begin{equation}
         e^{\beta} = \frac{1}{D}[u^2 + \frac{A^4}{36 \pi^2
         u^2}-\frac{A^2}{16\pi}]
         \label{Eq27}
          \end{equation}

  Where
  \begin{equation}
         u = \sqrt[3]{\frac{3}{2}D^2t+\sqrt{\frac{9}{4}D^4t^2 + \frac{A^6}{54
         \pi^3}}}
         \label{Eq28}
          \end{equation}

            [ Here we neglect the integration constant $t_0$ . ]\\

  In view of (27) one can easily find out  the other parameters as
\begin{equation}
         R^3 = \sqrt[\frac{3}{2}] {\frac{1}{D}[u^2 + \frac{A^4}{36 \pi^2
         u^2}-\frac{A^2}{16\pi}]}
         \label{Eq29}
          \end{equation}

\begin{equation}
         \theta=  \frac{3}{2}\sqrt{D e^{-3\beta}- \frac{A^2}{6 \pi}e^{-4\beta}}
         \label{Eq30}
          \end{equation}

 \begin{equation}
         \sigma^2= \frac{23}{12}[D e^{-3\beta}- \frac{A^2}{6
         \pi}e^{-4\beta}]
         \label{Eq31}
          \end{equation}

 \begin{equation}
         \rho =   \frac{A^2D^2}{12p(1-m)}\frac{1}{[u^2 + \frac{A^4}{36 \pi^2
         u^2}-\frac{A^2}{16\pi}]^2}
         \label{Eq32}
          \end{equation}
  \begin{equation}
         \frac{3}{4}\beta_0^2 =
         \frac{A^2(3m-1)}{24p(1-m)}e^{-4\beta}-\frac{D}{(1-m)}e^{-3\beta}
         \label{Eq33}
          \end{equation}

\bigskip
 \medskip

\cen{ \bf Behavior of the model : }

\medskip

 In this case we get the solution of matter field Universe . Here we note that our space
 time is singularity free . It has a particle horizon because\\

\cen{ $  \int_{t_0}^t \sqrt{\frac{D}{u^2 +\frac{A^4}{36\pi^2u^2}-
\frac{A^2}{6\pi}}} dt^\prime $}

               is a convergent integral .\\
The solutions represent an expanding model of the Universe .
    At $t \rightarrow \infty$ the expansion ceases . The gauge function decays during
    its evolution.\\

\bigskip
 \medskip

 \cen{ \bf 4.  Concluding Remarks: }

 \bigskip
 \medskip

    In this work, we have discussed Bianchi - I  cosmological model in Lyra geometry
considering a source free magnetic field , the reasons for which
has been discussed in the introduction. We have obtained three
sets of solutions with a special choice of metric coefficients  ,
viz.,  $\alpha  = a \beta$ . For a particular value of a  say,
$a= - 1 $, we observe that our model represents a singularity free
empty Universe . In the absence of the magnetic field  our model
reduces to the model as obtained by Singh and Singh (the model
has a initial singularity)[J.Math.Phys. 32, 2956, (1991)] . But
the nature of the solution are changed
 due to the presence of the magnetic field  ( here the model is singularity free).
The other choice of a say, $a = -\frac{1}{2}$  ,  we get a
solution of the matter filled Universe.


\end{document}